\documentclass[twocolumn,amsmath,amssymb]{snp}
\usepackage{graphicx}
\usepackage{dcolumn}
\usepackage{bm}
\usepackage{slashed}
\newcommand{\nn}{\nonumber}
\newcommand{\be}{\begin{equation}}
\newcommand{\ee}{\end{equation}}
\newcommand{\bea}{\begin{eqnarray}}
\newcommand{\eea}{\end{eqnarray}}

\newcommand{\om}{\omega}  
\newcommand{\vp}{\vec p}

\begin{document}

\title{{\Large  NJL model estimation of anisotropic electrical conductivity for
quark matter in presence of magnetic field}}

\author{\large Jayanta Dey}
\email{jayantad@iitbhilai.ac.in}
\affiliation{Indian Institute of Technology Bhilai, GEC Campus, Sejbahar, Raipur 492015, 
Chhattisgarh, India}
\author{\large Aritra Bandyopadhyay}
\affiliation{Departamento de F\'{i}sica, Universidade Federal de Santa Maria, Santa Maria, 
 RS, 97105-900, Brazil}
\author{\large Sabyasachi Ghosh} 
\affiliation{Indian Institute of Technology Bhilai, GEC Campus, Sejbahar, Raipur 492015, 
Chhattisgarh, India}
\author{\large Ricardo L. S. Farias}
\affiliation{Departamento de F\'{i}sica, Universidade Federal de Santa Maria, Santa Maria, 
 RS, 97105-900, Brazil}
\author{\large Gast\~ao Krein}
\affiliation{Instituto de F\'{i}sica Te\'orica, Universidade Estadual Paulista, 
01140-070 S\~ao Paulo, SP, Brazil} 
\maketitle

Present work has gone through the microscopic calculation of electrical conductivity of quark
matter in presence of magnetic field, where NJL model is considered for mapping the interaction
picture of the medium.
Let us start with Ohm's law'
\be
J_D^i = \sigma^{ij}E_j
\ee
where $J_D^i$ is dissipative current, $\sigma^{ij}$ electricl conductivity tensor and $E_j$ is electric field.
Now for a fluid of quark having spin-color degenracy $g$ and electric charge $q_f$ dissipative current 
from kinetic theory framework can be written as,
\be
J_D^j = q_f g \int \frac{d^3p}{(2\pi)^3} \vec v \: \delta f
\ee
Where $\delta f$ is small deviation of quark distribution function 
from the equilibrium Fermi-Dirac distribution of quark $f_0=\frac{1}{e^{\beta\om}+1}$. 
Interms of 3-momentum $(\vec p)$ and energy $(\om)$ particle velocity can be writen as $\vec v = \frac{\vec p}{\om}$.
Now to find $\delta f$ in presence of electric field $\vec E$ and magnetic field $\vec B$ we 
use RTA in Boltzmann's equation, where we can assume the a general force term
\be
\vec{\cal F} = \alpha \vec{e} + \beta \vec{b} 
+ \gamma \vec{e}\times \vec{b}
\ee 
where $\vec e$, $\vec b$ are unit vector along $\vec E$ and $\vec B$.
With suitable connection between $\delta f$ and $\vec{\cal F}$~\cite{Sedrakian,JD_QGPB},
the coefficients $\alpha$, $\beta$ and $\gamma$ can be found as
\bea
\alpha &=& q \left(\frac{\tau_c}{\om}\right)\frac{1}{1+(\tau_c/\tau_B)^2}\, E,
\label{alpha}
\nn\\
\beta &=& q \left(\frac{\tau_c}{\om}\right)\frac{(\tau_c/\tau_B)^2}{1+(\tau_c/\tau_B)^2}
\, (\vec {e}\cdot \vec{h}) E,
\label{beta}
\nn\\
\gamma &=& - q \left(\frac{\tau_c}{\om}\right) \frac{(\tau_c/\tau_B)}{1+(\tau_c/\tau_B)^2}
\, E,
\label{gamma}
\eea
where $\tau_B$ and $\tau_c$ as magnetic and thermal relaxation time.
After taking care of all degenracy factors of $u$ and $d$ quarks, we get 3
components of electrical conductivity, whose general expressions can be written as 
\be
\sigma_n =  e^2 {\beta}\; \frac{20}{9}  \int \frac{d^3p}{(2\pi)^3}\, 
\frac{{\vp}^2}{\om^2} 
\frac{\tau_c(\tau_c/\tau_B)^n}{1+(\tau_c/\tau_B)^2} f_0(1-f_0).
\label{cond_ne}
\ee
with $n=0,1,2$. We will get temperature ($T$) and magnetic field ($B$) dependent
quark mass from NJL model, briefly discussed below, which will be used in above Eq.~(\ref{cond_ne})
to estimate $\sigma_n$ of quark matter.
\begin{figure}
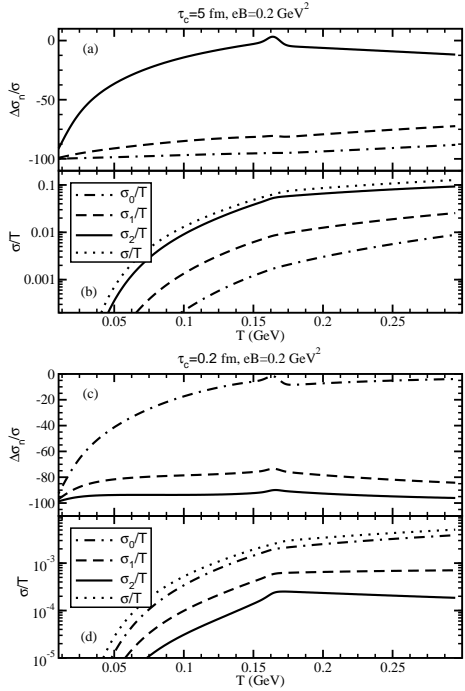

	\begin{center}
		\includegraphics[scale=0.25]{el_T1.eps}
		\includegraphics[scale=0.25]{el_T2.eps}
		\caption{Temperature dependence of three components of electrical 
			conductivities ($\sigma_{0,1,2}$) with $eB=0.2$ GeV$^2$ and $\sigma$ (without $B$) for 
			$\tau_c=5$ fm (b) and $0.2$ fm (d). $\Delta\sigma_n/\sigma=(\sigma_n -\sigma)/\sigma$
			for $\tau_c=5$ fm (a) and $0.2$ fm (c).} 
		\label{el_T}
	\end{center}
\end{figure}

The Lagrangian density for the isospin-symmetric two-flavor version of NJL model 
in presence of electromagnetic field ($A^\mu$) is given by
\bea
\nn \mathcal{L}_{NJL}= -\frac{1}{4} F^{\mu\nu}F_{\mu\nu}
+ \bar{\psi}\left(\slashed{D}-m\right)\psi
\\ + G\left[ (\bar{\psi}\psi)^2
+(\bar{\psi}i\gamma_5{\vec\tau}\psi)^2\right],
\label{NJL_lag}
\eea
In quasi-particle approximation, the gap equation for the constituent quark mass $M$ 
at finite $T$ and $B$ is given by
\be
M = m - 2 G  \sum_{f=u,d}\langle \bar{\psi}_f\psi_f\rangle,
\label{Gap_B}
\ee
where $\langle \bar{\psi}_f\psi_f\rangle$ represents the quark condensate of flavor~$f$, and
a thermo-magnetic NJL coupling 
constant $G(B,T)$ has been considered~\cite{Ricardo,Ricardo2}.
%
%

Using the governing NJL model Eq.~(\ref{Gap_B}) and then plug in to Eq.~(\ref{cond_ne}),
we have estimated $\sigma_n$. 
In Figs.~\ref{el_T}(b) and (d),
we present the temperature dependence of the different components of electrical 
conductivities $\sigma_n$ (scaled with $T$) in presence of an external magnetic field as well as without 
field case,
\be
\sigma =  e^2 {\beta}\; \frac{20}{9}  \int \frac{d^3p}{(2\pi)^3}\, 
\frac{{\vp}^2}{\om^2} \tau_c f_0(1-f_0).
\label{cond_B0}
\ee
Their difference $\frac{\Delta\sigma_n}{\sigma}=\frac{(\sigma_n-\sigma)}{\sigma}$
are plotted in Figs.~\ref{el_T}(a) and (c). 
The results are presented for $\tau_c=5$ fm (a, b) and $\tau_c=0.2$ fm (c, d) but in same magnetic
field ($eB=0.2$ GeV$^2$), which can be
assigned with the zones $\tau_c>\tau_B$ and $\tau_c<\tau_B$ respectively.
It means that $eB=0.2$ GeV$^2$ may be considered as stronger magnetic field for $\tau_c=5$ fm
and weaker magnetic field for $\tau_c=0.2$ fm. Therefore, former case is showing $\sigma_2>\sigma_0$
and latter case is showing $\sigma_2<\sigma_0$. It is controling by the anisotropic 
function $\frac{(\tau_c/\tau_B)^n}{1+(\tau_c/\tau_B)^2}$. In terms of anisotropy, the above
outcomes can be breifly sketched as
\bea
\nn {\rm for}~ \tau_c&=&5 ~{\rm fm} \\
\sigma^{xx}&=&\sigma^{yy}<\sigma^{zz}\Rightarrow {\rm larger~ anisotropy} 
\\
\nn {\rm for}~ \tau_c&=&0.2~ {\rm fm}~,~\\ 
\sigma^{xx}&=&\sigma^{yy}\approx\sigma^{zz}\Rightarrow {\rm smaller~ anisotropy}~~~~~, 
\label{anisotropy_eq}
\eea
when external magnetic field $eB=0.2$ GeV$^2$ is in the z-direction.
As seen from Fig.~\ref{el_T}, 
all the components of the electrical conductivities increase with temperature with a kink near 
the quark-hadron phase transition temperature $T_c$ and rate of increments are also different 
for two different phases.

\end{document}